\documentclass[times]{article} 
\usepackage{graphicx,times,amsmath, amssymb,subfigure}
\usepackage[T1]{fontenc} 
\usepackage{fixltx2e} 
\usepackage{xspace}

\begin{document}
\title{A Topological Investigation of Phase Transitions of Cascading Failures in Power Grids}
\author{Yakup Ko\c{c}$^{1,2,}$\thanks{Corresponding Author,~~Email: \texttt{Y.Koc@tudelft.nl},  Address: Jaffalaan 5, 2628BX Delft, The Netherlands, Phone: +31 (0)15 27 88380}~~Martijn Warnier$^1$~~Piet Van Mieghem$^{2}$~~\\
Robert E. Kooij$^{2,3}$~~Frances M.T. Brazier$^1$\\
\\
 $^1$ Faculty of Technology, Policy and Management,\\
  Delft University of Technology, the Netherlands\\
 $^2$ Faculty of Electrical Engineering, Mathematics and Computer Science,\\
  Delft University of Technology, the Netherlands\\
 $^3$TNO (Netherlands Organisation for Applied Scientific Research), the Netherlands
 }
\date{}

\maketitle
\thispagestyle{empty}

\begin{abstract}
Cascading failures are one of the main reasons for blackouts in electric power transmission grids. The economic cost of such failures is in the order of tens of billion dollars annually. The loading level of power system is a key aspect to determine the amount of the damage caused by cascading failures. Existing studies show that the blackout size exhibits phase transitions as the loading level increases. This paper investigates the impact of the topology of a power grid on phase transitions in its robustness. Three spectral graph metrics are considered: spectral radius, effective graph resistance and algebraic connectivity. Experimental results from a model of cascading failures in power grids on the IEEE power systems 
demonstrate the applicability of these metrics to design/optimise a power grid topology for an enhanced phase transition behaviour of the system.   
  
\end{abstract}

\section{Introduction}
\label{sec_Introduction}
	The electric power grid is of key importance to modern societies, not only because of the vital role of electric power for daily life, but also because of the strong dependencies of other critical infrastructures (such as telecommunications, transportation, and water supply~\cite{Eeten2011, Buldyrev2010}) on power grids. Security and continuous availability of power supply is crucial. Disruption of power delivery systems potentially causes severe effects on the public order and substantial economic cost for the society~\cite{Hines2009}. Although secure and safe operation makes the power grid one of the most robust critical infrastructures and significantly reduces the risk of large-scale blackouts, many countries still suffer from catastrophic blackouts paralysing the daily life~\cite{BrazilBlackoutRef, blackoutReport}. 

	A cascading failure is an initial disturbance followed by a sequence of dependent failures of individual components. Each failure successively weakens and stresses the system as a whole making the subsequent failures more likely~\cite{Baldick2009}. The networked structure of complex power grids makes it possible to analyse and better understand interdependencies between its components by adopting a Complex Networks approach. In such an approach, the collective emergent behaviour of complex power grids is assessed instead of the detailed behaviour of a system at the individual components level.

	
	Interconnectedness of the power grid enables shipping of bulk electric power over the grid. However, it also allows the spread of local failures into the global grid resulting into large-scale cascading failures. In a power grid, a single failure may propagate into the rest of the network in different ways including instabilities in frequency and voltage levels, human mistakes, hidden failures in the protection mechanism, and line overloads. Interdependencies between components become stronger as the loading level of the system increases, and consequently, more significant damage emerges due to the cascade. For example, a failure of a transmission line results in much larger damage due to cascading overloads if the system is over stressed. Although a higher loading level almost always results in a larger blackout size, the relationship between loading level and blackout size is not linear. Previous studies~\cite{Pahwa2014, Dobson2007_2, Nedic2006, Liao2004} have shown that the blackout size exhibits phase transitions when increasing loading level. This suggests the existence of critical loading points at which large and abrupt change occurs in the size of the damage due to cascades with only small changes in the network loading. 
	
	
	
	 Although many researchers~\cite{Pahwa2014, Dobson2007_2, Nedic2006, Liao2004, Wang2011,Koc2013_3} have investigated/reported the existence of phase transitions in the robustness of power grids, to the best of our knowledge, no attention has been paid to the assessment of the impact of a power grid topology on the phase transitions in its robustness. Phase transitions occur in many other complex systems such as e.g. in network synchronisation and epidemic processes~\cite{Wang2003, Mieghem2009, Restrepo2005, Zhu2013, Louzada2013}. In these fields/studies, the phase transition phenomena is analysed from a complex networks point of view, and related to spectral graph properties of the corresponding graph of the system. Inspired by these studies, this paper investigates the impact of the topology on the phase transitions in the robustness of a power \emph{transmission} grid by focussing on three spectral graph metrics~\cite{Mieghem2011}: spectral radius, algebraic connectivity and effective graph resistance.

\section{A model of cascading failures in power grids}
\label{sec_Model}

A power grid is a three-layered complex interconnected network consisting of generation, transmission, and distribution parts. Electric power is shipped from the generation buses to distribution substations through the transmission buses, all interconnected by transmission lines. The impedances, voltage levels at each individual power station, voltage phase differences between power stations and loads at terminal stations control power flow in the grid. An ideal model of power grids necessarily captures the continuous dynamics of generators, discrete dynamics and hidden anomalies in protection mechanisms, non-linear AC equations that govern the power flow throughout the network, and the behaviour of the operators~\cite{Hines2010}. To reduce complexity, each model deploys a certain level of abstraction to mimic the power grids.

Due to strong couplings between components, a local disturbance affects not only the neighbouring components, but also the far away components in a power grid. Models relying on Kirchoff Laws (e.g.~\cite{Pahwa2014, Carreras2002} ) capture this interdependency between power system components, unlike models deploying topological approaches in which the excess power is distributed over neighbouring components based on local redistribution rules~\cite{Crucitti2004, Wang09, Wang2011, Sole2008}. This paper deploys a similar model to~\cite{Pahwa2014, Carreras2002} relying on Kirchoff Laws, introduced by Ko\c{c} et al.~\cite{Koc2013_3} and implemented in MATCASC, a MATLAB based tool to simulate cascading line overloads~\cite{Koc2013_3}.

MATCASC models a power transmission grid as a graph to analyse the cascading effect due to line overloads (See Figure~\ref{fig:IEEE30}). In a graph representation of a power grid, nodes represent generation, transmission, distribution buses, substations and transformers, while links model the transmission lines. The links are weighted by the admittance (or impedance) value of the corresponding transmission lines.

In line with many other researchers~\cite{Dobson2007, Bao2009, Kinney2004}, MATCASC computes the flow values throughout a network by DC load flow analysis. An initial failure of a component changes the balance of power flow distribution over the grid and causes redistribution of power flow over the network. This dynamic response of the system to this triggering event might overload other parts in the network. The protection mechanism trips these newly overloaded components, and the power flow is again redistributed potentially resulting in new overloads. In a power grid, each line has a relay protecting it from permanent damage due to e.g.\xspace excessive flows. If overloaded, an over-current relay notifies a circuit breaker to trip a line when the current of the line exceeds its rated limit and this violation lasts long enough to permanently damage the line. The rated limit of a transmission line is assumed to be proportional to it's initial flow with a tolerance level~\cite{Motter2002, Motter2004, Crucitti2004, Crucitti2004_2, Crucitti2005} $\alpha$. For the sake of simplicity, this paper assumes a deterministic model for the line tripping mechanism, i.e.\xspace a circuit breaker for a line trips at the moment the flow of the line exceeds its rated limit. In case of islanding, cascading failures continue in each island in which generators or loads are shed respectively to attain a supply-demand balance. The cascade of failures continues until no more components are overloaded. After the cascade subsides, the robustness of the grid against cascading failures is quantified in terms of the fraction of the served power demand after the cascading failures.

The cascading failure simulation process is described as follows:

\begin{enumerate}
\item Select a line threat either randomly or based on a specific attack strategy.
\item Remove the line threat, update the grid topology (i.e. admittance matrix~\cite{Schavemaker2008}), and compute the line flow values based on DC load flow analysis.
\item Check the connectedness of the grid. If there are islands, in each island the cascading failures continue separately. In each island, adjust generation and load respectively to attain a power balance.
\item Check the flow limit violations of the transmission lines. If the flow value of a transmission line exceeds its rated limit, label the corresponding line as the line threat, and repeat steps 2, 3, and 4.
\item If there are no further line overloads, report the shed/still satisfied (served) power demand after the cascade. 
\end{enumerate}

\section{Spectral Graph Metrics}
\label{sec_Metrics}

This section elaborates on complex network preliminaries including definitions of relevant concepts, and relevant metrics from spectral graph theory including spectral radius, algebraic connectivity, and effective graph resistance.

A network \emph{G($\mathcal{N}$,$\mathcal{L}$)} consisting of a set $\mathcal{N}$ of \emph{N} nodes and a set of $\mathcal{L}$ of \emph{L} links, can be represented by its \emph{adjacency matrix} $A$. The adjacency matrix of a simple, unweighted graph \emph{G($\mathcal{N}$,$\mathcal{L}$)} is an $N \times N$ symmetric matrix reflecting the interconnection of the nodes in the graph: $a_{ij}=0$ indicates that there is no edge, otherwise $a_{ij}=1$. For a weighted graph, the network is represented by a weighted adjacency matrix $W$ where $w_{ij}$ corresponds to the weight of the link between nodes $i$ and $j$; a weight can be e.g. a distance, cost, or delay. 

The eigenvalues of an adjacency matrix of a graph reveal important characteristics of a network. The largest eigenvalue of a graph with adjacency matrix $W$ is called~\cite{Mieghem2011} the \emph{Spectral radius} $\rho(W)$ of the network. The spectral radius plays an important role in dynamic processes on networks. For instance, in epidemics, a phase transition occurs at the epidemic threshold  $\tau_{c}$ for virus spreading rate $\tau$: if $\tau$ is smaller than $\tau_{c}$, then the virus dies out, whereas for $\tau$ larger than $\tau_{c}$, the virus spreads over the network. Van Mieghem et al.~\cite{Mieghem2009} relate the spectral radius of a network directly to its robustness with respect to virus spread in the network, and shows that the epidemic threshold is the inverse of the spectral radius of the network. In addition to the virus spread, the same type of phase transition threshold in the coupling strength $g_{c} \sim \frac{1}{\rho}$
occurs in a network of coupled oscillators~\cite{Restrepo2005}.


The weighted \emph{Laplacian matrix}~\cite{Mieghem2011} $Q$ is another way to fully characterize a graph, and is defined as:
\begin{equation}\label{Laplacian}
Q=\Delta - W 
\end{equation}

\noindent where $\Delta$ is the diagonal matrix of the strengths of $G$: $\delta_{i}$=$\sum_{j}^{N}{w_{ij}}$. Hence, the Laplacian can be constructed as follows: 
\begin{equation}
  Q_{ij}=\begin{cases}
      
    \delta_{i}, 	& \text{if $i=j$}.\\
     -w_{ij},			& \text{if $i \neq j$ and $(i,j) \in L$}\\
     0,				& \text{otherwise}.

  \end{cases}
\end{equation}

The Laplacian matrix constructed by susceptance (inverse of reactance~\cite{Schavemaker2008}) values is equivalent to admittance matrix in power systems theory. Because the Laplacian is symmetric, positive semi-definite and its rows sum up to zero, the eigenvalues of the Laplacian are positive and real except for the smallest one $\mu_{N}$ that is zero. The \emph{algebraic connectivity}~\cite{Mieghem2011} $\mu_{N-1}$ of a network is the second smallest eigenvalue of its Laplacian matrix. Algebraic connectivity reflects how well a network is topologically connected. A relatively large $\mu_{N-1}$ indicates a relatively well-connected topology while a zero value for $\mu_{N-1}$ means that the network is disconnected.

The effective resistance~\cite{Mieghem2011} $R_{ij}$ between a pair of nodes $i$ and $j$ is the potential difference between these nodes when a unit current is injected at node $i$ and withdrawn at node $j$. The effective graph resistance $R_{G}$ is the sum of the individual effective resistances between each pair of nodes in the network. The effective graph resistance is a measure of how well a graph is connected in terms of electrical resistance. The existence of parallel paths between two nodes in a physical power grid topology, and a homogeneous distribution of their impedance values result in a smaller effective resistance between these two nodes (i.e.\xspace a stronger electrical connection between these nodes). Ko\c{c} et al.~\cite{Koc2013_4, Koc2013_5} show that the effective graph resistance relates the topology of a power grid to its robustness against cascading failures. A power grid topology with relatively small $R_{G}$ results in a higher level of robustness against cascading failures by targeted attacks. The effective graph resistance of a network also equals the sum of the inverse of the eigenvalues of its Laplacian matrix:

\begin{equation}\label{EffGraphResistanceEigenValues}
R_{G}=N{\sum_{i=1}^{N-1}} \frac{1}{\mu_{i}}
\end{equation}

 \noindent where $\mu_{i}$ is the $i^{th}$ eigenvalue of the Laplacian matrix, and $\mu_{1} \geq \mu_{2} \geq ...\geq \mu_{N-1} \geq \mu_{N}$.

\section{Experimental Methodology}
\label{sec_Experimental Methodology}

The experimental method used in this analysis is to consider a variety of power systems and subject them to deliberate attacks under various loading levels to determine the phase transition. This approach makes it possible to assess the relationship between the spectral graph metrics and phase transition behaviour of the power grid. This section elaborates on the deployed power systems, attack strategies, and determination of the phase transition point.

\subsection{Power grids: IEEE power systems and the synthetic grids}
\label{subsec_Test systems}
 
 The computation of the spectral graph metrics for a power grid requires data describing its topology (i.e.\xspace interconnection of buses with lines) and  admittance values of transmission lines, while simulation of cascading failures to determine the phase transition behaviour necessitates information about the number of buses, their types and finally their generation capacity and load values. As IEEE power systems provide all of these data, they are considered as the reference systems. Due to the limited number of the IEEE power systems, additional synthetic power systems are created based on these reference IEEE power systems.
 
 The synthetic power systems generated for the purpose of experimentation have exactly the same properties as the reference IEEE test system (e.g.\xspace topology, number of buses and links, type of buses and their demand-generation capacity values) except for the admittance matrix. To create a new synthetic power system based on the IEEE reference system, to manipulate the spectral graph metrics and the phase transition behaviour, e.g.\xspace 10\% of the lines of the reference power system are randomly chosen and impedance values of these transmission lines are increased by themselves. For example, when creating a synthetic power system based on the IEEE 30 power system~\cite{TestCaseRef} (consisting of 30 buses and 41 lines), 4 transmission lines are randomly chosen: $l_{1}$, $l_{2}$, $l_{3}$, $l_{4}$. Then the impedance values of these lines (i.e. $x_{1}$, $x_{2}$, $x_{3}$, $x_{4}$) are doubled so that a new synthetic power system is created with its lines $l_{1}$, $l_{2}$, $l_{3}$, $l_{4}$ having the new impedance values: $2x_{1}$, $2x_{2}$, $2x_{3}$, $2x_{4}$. This change of impedance values results in a different power flow distribution over the network, and subsequently a different level of robustness for each network. A second synthetic test system is subsequently created from the reference IEEE 30 power systems by multiplying the impedance values of these lines by a factor of three so that $l_{1}$, $l_{2}$, $l_{3}$, $l_{4}$ have the impedance values of $3x_{1}$, $3x_{2}$, $3x_{3}$, $3x_{4}$, respectively. 

To create the synthetic power grids, the impedance values of transmission lines are manipulated rather than removing/adding lines to the real(istic) IEEE power system topology. This approach preserves the realistic structure of the systems while it generates a large number of synthetic power systems.

\subsection{Attack Strategies}
\label{subsubsec_Attack Strategies}

To determine phase transition behaviour, the response of a power system to attacks is assessed under various loading levels. This paper designs attack strategies based on (i) edge betweenness and (ii) electrical node significance centralities.

The edge betweenness centrality~\cite{Mieghem2006} of a link is a graph topological metric quantifying the topological centrality of a link in a network. The edge betweenness centrality of a link \textit{l} is defined as the normalized number of shortest paths between any pair of nodes passing through \textit{l}.

\begin{equation}\label{Betweenness}
C_{B}(l)=\sum_{\substack{s,t \in N }}{\frac{\sigma_{st}(l)}{\sigma_{st}}}
\end{equation}

\noindent where $\sigma_{st}(l)$ is the number of shortest paths passing through line \textit{l}, while $\sigma_{st}$ is the total number of shortest paths in the grid topology. An attack based on betweenness centrality targets the line with the highest betweenness centrality.

The electrical node significance~\cite{Koc2013, Koc2013_2} is a contextual node centrality measure, specifically designed for power grids. The electrical node significance $\delta$ of a node \emph{i} is defined as the amount of power distributed by node \emph{i}, normalized by the total amount of the power that is distributed in the entire grid:

\begin{equation}\label{Delta}
\delta _{i}=\frac{P_{i}}{\sum_{j=1}^{N} P_{j}}
\end{equation}

\noindent where $P_{i}$ stands for the total power distributed by node $i$. An attack based on $\delta$ requires targeting \emph{the most heavily loaded outgoing link from the node with the highest electrical node significance} in the network. Removal of this link most likely results in the largest cascading failure in the power network~\cite{Verma2012}.

\subsection{Computation of the critical loading threshold}
\label{subsec_Experimental  values}

An effective way of determining the phase transition point of a power grid is to assess the network robustness for a broad range of network loading levels $l$. The loading level of a component is the ratio of the actual load to the rated limit (i.e. maximum capacity) of the component, and it is assumed to be uniform throughout the network. An increase/decrease of $l$ is achieved by simultaneously increasing/decreasing the power demand by a factor while keeping the network fixed. 
The damage resulting from cascading failures depends on the network loading level $l$ of the system. When the value of $l$ is sufficiently large, an arbitrary attack causes grid to collapse because the capacity of each line is strictly limited. On the other hand, if $l$ is sufficiently small, because all the lines have larger margins to afford additional power flow, no substantial damage emerges and the grid remains functional. With the increase of $l$, a crossover behaviour of the system occurs from no breakdown to a large scale breakdown. This crossover behaviour is represented by a critical loading threshold $l_{c}$, at which a transition occurs from the normal state to the collapse. When $l<l_{c}$, the grid maintains its normal state after an attack, while when $l>l_{c}$, an attack collapses the grid because the lines do not have enough capacity to afford the excess power.

\begin{figure*}[!htb]
\centering
\includegraphics[scale=0.45]{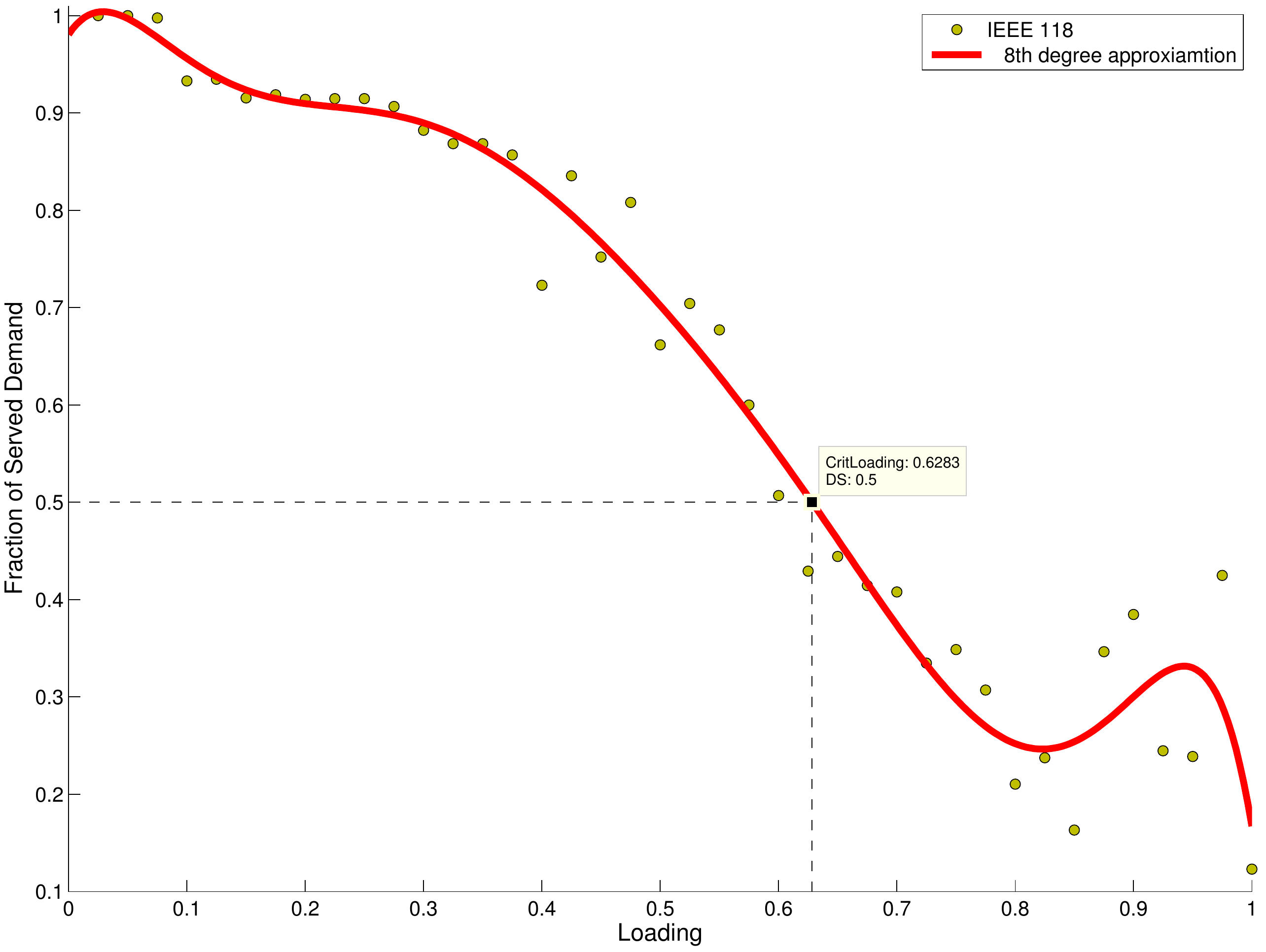}
\caption{The robustness curve, its 8$^{th}$ order polynomial approximation, and the corresponding critical loading threshold for IEEE 118 power system}
\label{fig:ApprIEEE118RobCurve}
\end{figure*}

Figure~\ref{fig:ApprIEEE118RobCurve} plots the fraction of served power demand after cascades (induced by electrical node significance based attacks) as a function of the network loading level for the IEEE 118 power system (consisting of 118 buses and 181 lines). Figure~\ref{fig:ApprIEEE118RobCurve} shows that for the region of $l<0.4$ and for $l>0.8$, the network damage is stable. On the other hand, a transition from small-scale to large-scale damage occurs in $0.8>l>0.4$. Accordingly, a small increase in $l$ disproportionally increases the damage caused by an attack on a network.

In complex systems and statistical physics, a phase transition entails an abrupt change in some measured quantities of the system when the system goes through a critical point. For a type-1 phase transition, this change is discontinuous, while for a type-2 transition, the change is continuous~\cite{Dobson2007_2}. In the context of power grids, when the measured quantity is the robustness of the system to cascades, a critical loading point can be determined accurately for type-1 transitions due to discontinuity in the robustness of the system. Such clear critical loading points do not exist for type-2 phase transitions. Figure~\ref{fig:ApprIEEE118RobCurve} suggests that a power system can exhibit a type-1 phase transition in its robustness. To enable determination of a critical point for each power system, this paper defines the critical loading threshold $l_{c}$ as the loading level at which an attack results in the loss of 50\% of the total power demand. To determine $l_{c}$, the robustness curve of a system is approximated by an $8^{th}$ order polynomial. Figure~\ref{fig:ApprIEEE118RobCurve} illustrates the concepts of robustness curve, its $8^{th}$ order polynomial approximation, and critical loading threshold for the IEEE 118 power system. It shows that the 50\% loss of served power demand occurs as a result of an attack on the grid when $l$=0.62, hence $l_{c}$=0.62. 

As observed in other recent studies~\cite{Pahwa2014, Ma2013}, Figure~\ref{fig:ApprIEEE118RobCurve} shows that the fraction of served power demand is, counter intuitively, a non-monotonic decreasing function of the loading level in the system. Consequently, an increase/decrease in the loading level may occasionally result in a more/less robust network. This suggests existence of Braess's Paradox in the electrical power grids, stating that adding/creating extra capacity to/in a network can occasionally reduce the overall performance of a network~\cite{Braess2005}.

\section{Numerical Analysis}
\label{sec_Experimental Verification}
This section investigates the impact of topological changes on the phase transition behaviour of a power grid, by assessing the relationship between critical loading threshold and spectral radius, algebraic connectivity, and effective graph resistance in a power grid. First the impact of line removal on the phase transition in power grids are investigated. Then, the relationship between the topology and the phase transitions is assessed quantitatively.

\subsection{Impact of line removal on phase transition in power grids}
\label{subsec_Impact of line removal}

\begin{figure*}
\begin{center}	
	\includegraphics[width=.45\textwidth]{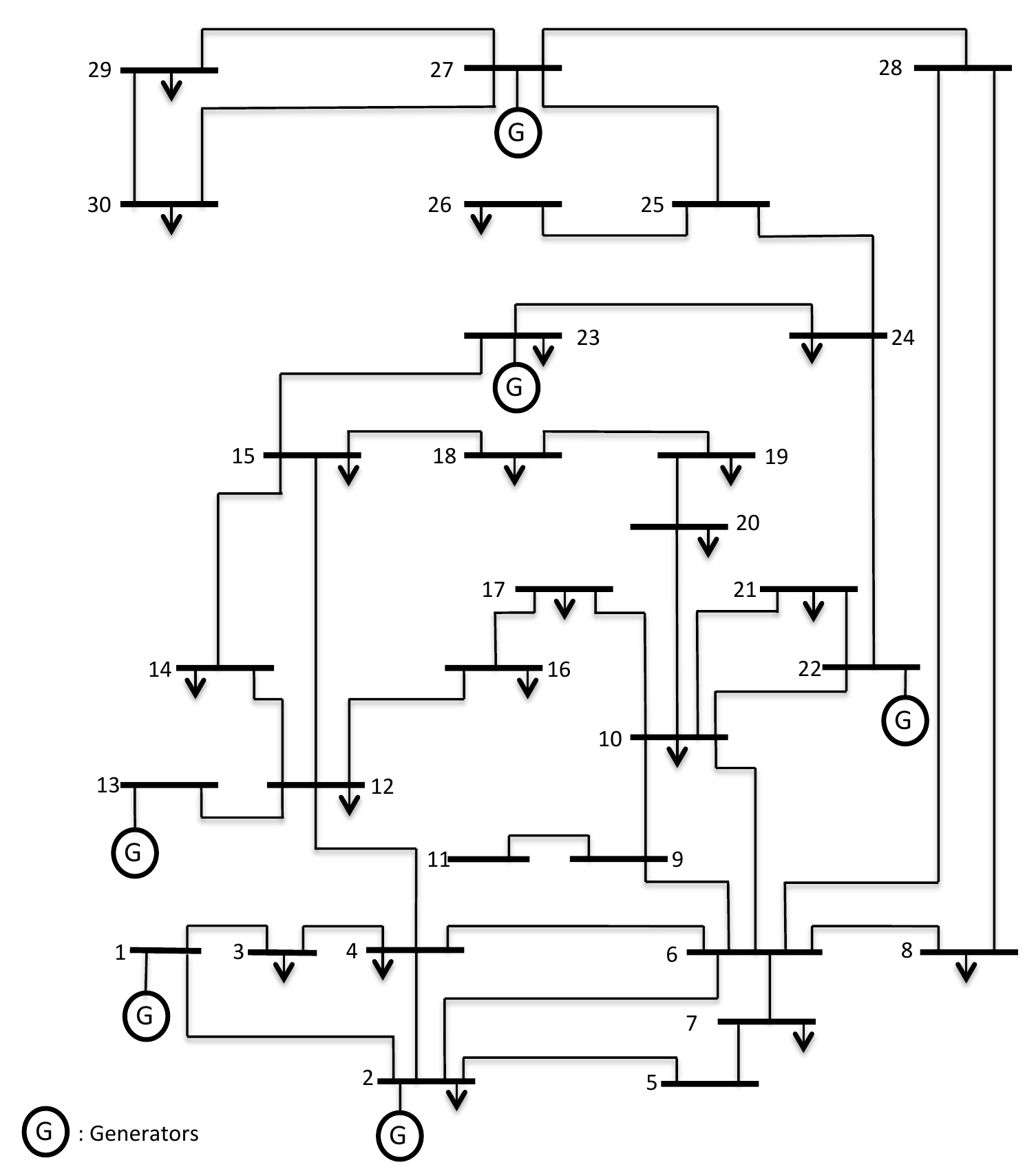}
	\label{fig:IEEE30SingleLine}
	\includegraphics[width=.52\textwidth]{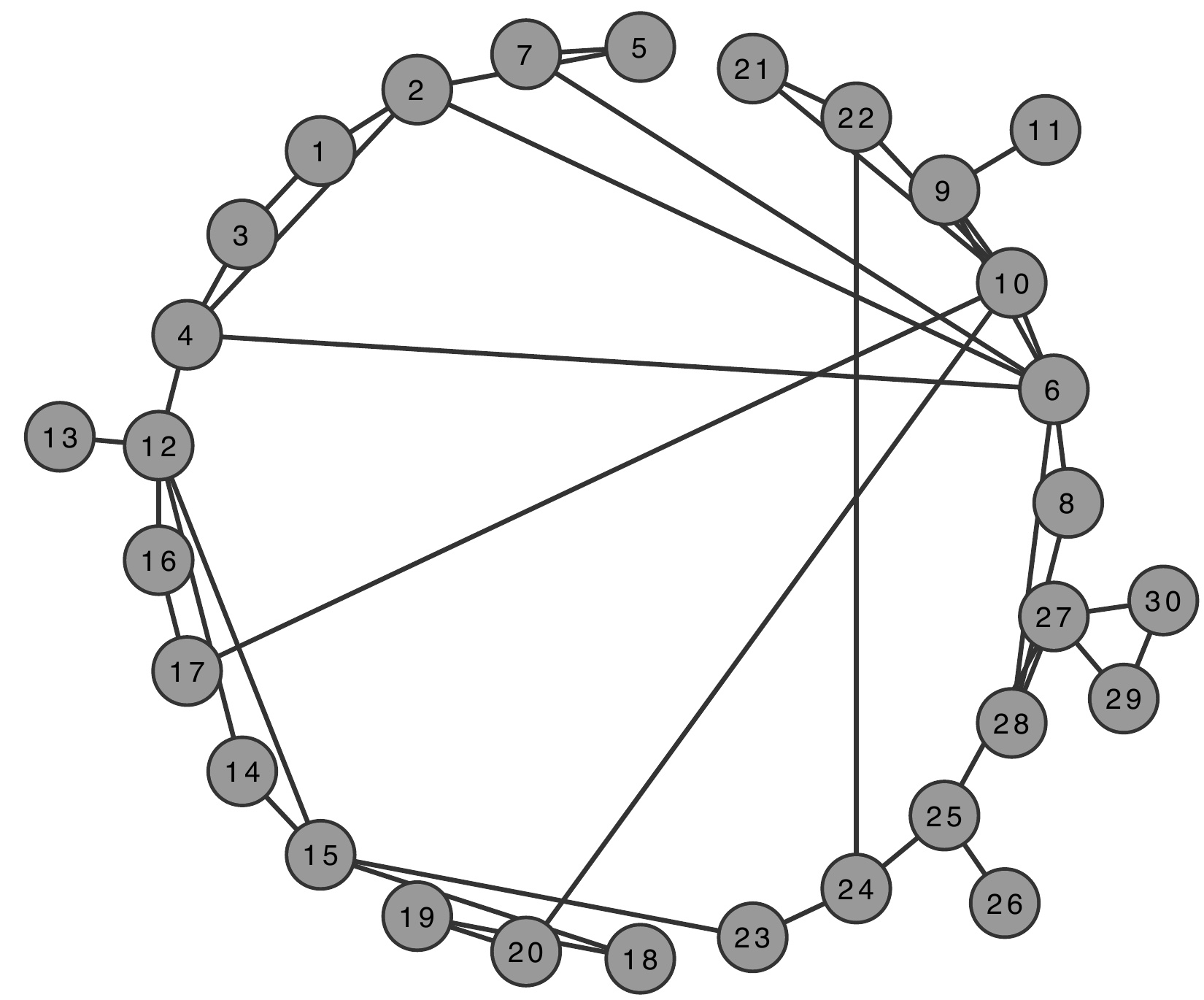}
	\label{fig:IEEE30Graph}
\caption{The single line diagram and the graph representation of IEEE 30 power system.}
\label{fig:IEEE30}
\end{center}
\end{figure*}

\begin{figure*}
    \centering
    \subfigure[]{\includegraphics[width=0.8\textwidth]{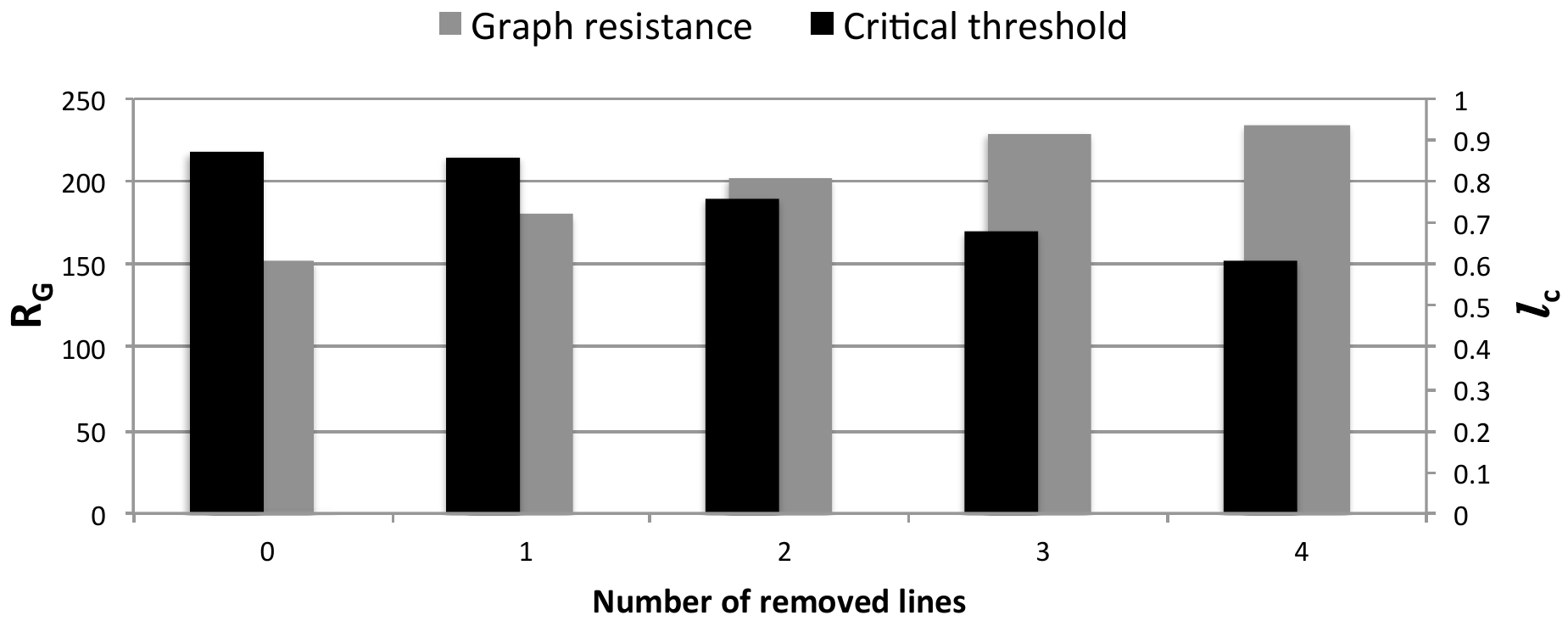}}
    \subfigure[]{\includegraphics[width=0.8\textwidth]{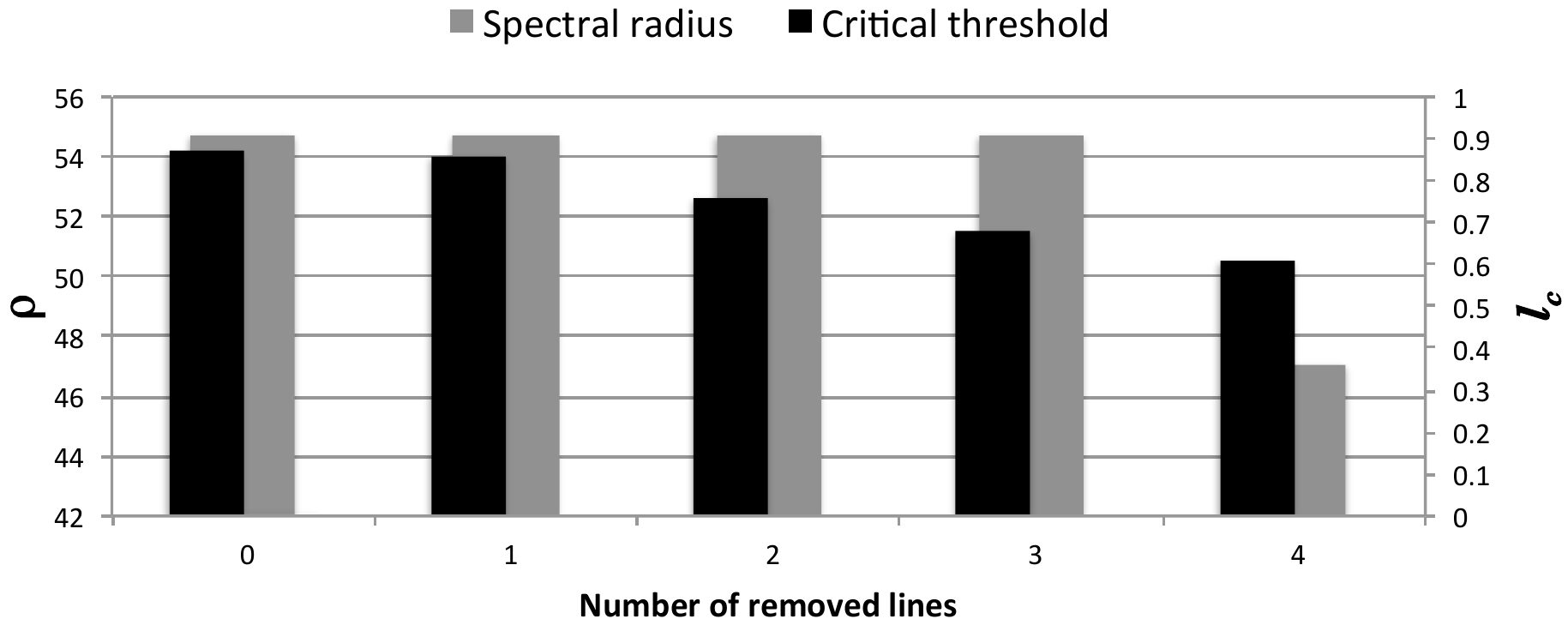}}
    \subfigure[]{\includegraphics[width=0.8\textwidth]{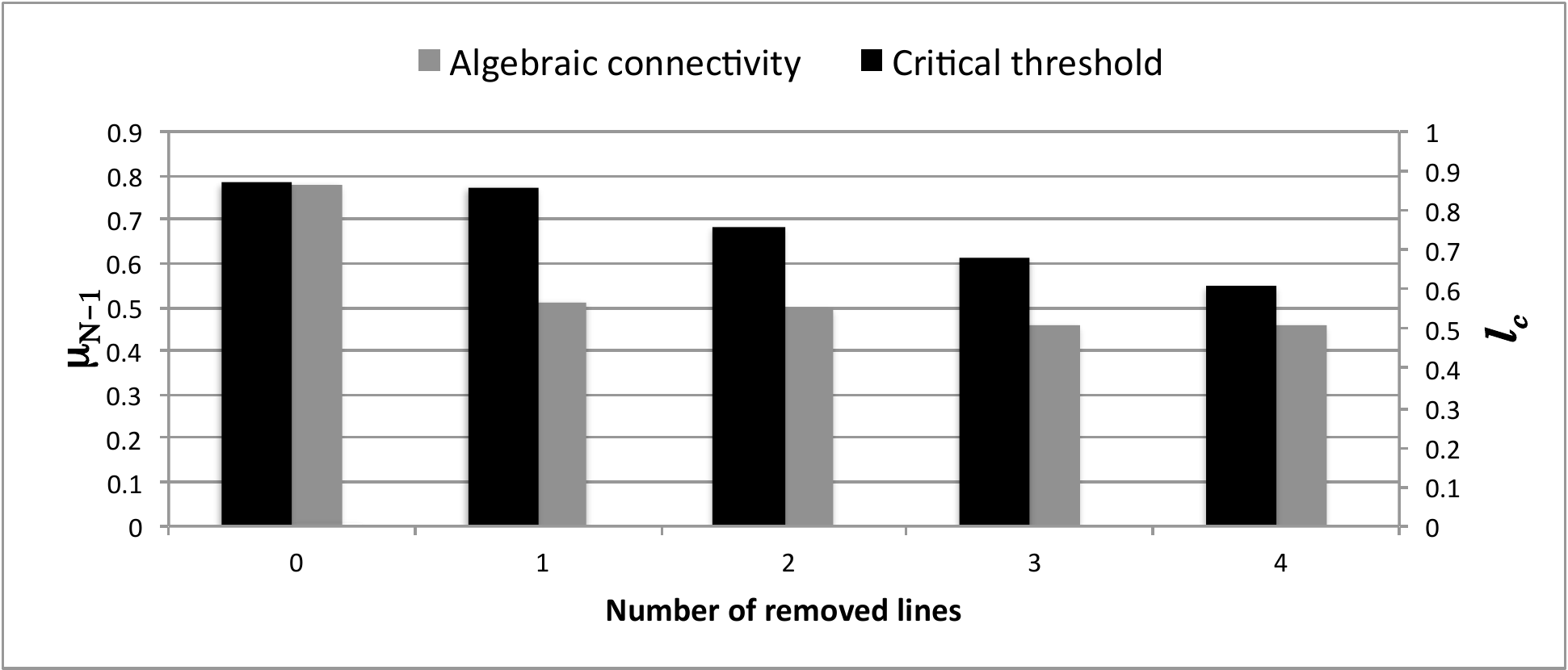}}

    \caption{The impact of line removal on  $l_{c}$ vs. $R_{G}$, $\rho$, and $\mu_{N-1}$ in the IEEE 30 power system. The grid is attacked based on betweenness centrality.}
\label{fig:IEEE30MetricsVsThreshold}
\end{figure*}

To assess the impact of line removal on the phase transitions in a power grid, the IEEE 30 power system, consisting of 30 buses and 41 lines (See Figure~\ref{fig:IEEE30}) is considered as a use case. 10\% of its lines (i.e. 4 lines) are randomly chosen: line ID 15 (connecting bus 4 to bus 12), line ID 24 (connecting bus 19 to bus 20), line ID 29 (connecting bus 21 to bus 22) and line ID 37 (connecting bus 27 to bus 29). The IEEE 30 power system is weakened by removing these randomly chosen lines progressively. At each time, the spectral graph metrics of the weakened system are computed and the critical loading threshold is determined by attacking the network based on betweenness centrality. Figure~\ref{fig:IEEE30MetricsVsThreshold} shows the result.

Figure~\ref{fig:IEEE30MetricsVsThreshold} shows how line removal affects the spectral graph metrics and the critical loading point of the IEEE 30 power system. As expected, removing a line from the topology of a power system weakens it and results in a smaller $l_{c}$, i.e. an earlier phase transition. For instance, in Figure~\ref{fig:IEEE30MetricsVsThreshold}.a, removing the line with ID 15 (connecting bus 4 to bus 12 in Figure~\ref{fig:IEEE30}) from the IEEE 30 power system topology shifts the critical loading point of the system from a loading level of 87\% to 86\%. As more lines are removed from the topology, $l_{c}$ becomes smaller and smaller with different gradients. $R_{G}$ and $\mu_{N-1}$ capture this behaviour of $l_{c}$ to different extents, while $\rho$ behaves differently than $l_{c}$ as more lines are removed from the topology. $R_{G}$ and $\mu_{N-1}$ have similar patterns; as lines are removed progressively, $R_{G}$ becomes smaller and smaller while $\mu_{N-1}$ becomes larger and larger. On the other hand, the $\rho$ of the network shows a different pattern; progressive line removal results first in a very slight, and then in a very steep change in the network spectral radius.  

In Figure~\ref{fig:IEEE30MetricsVsThreshold} , removing transmission lines from the IEEE 30 power system increases the electrical distances between the buses and results in an increased $R_{G}$. On the other hand, these line removals decreases the topological connectivity and diameter~\cite{Mieghem2009} of the network, therefore $\mu_{N-1}$ and $\rho$ decrease. These trends in the spectral graph metrics in Figure~\ref{fig:IEEE30MetricsVsThreshold} also comply with the fundamental characteristics of these metrics. $R_{G}$ is a monotonic non-increasing function of the number of transmission lines in a power grid network~\cite{Ellens2011}. Hence, in Figure~\ref{fig:IEEE30MetricsVsThreshold}, removing the transmission lines from the IEEE 30 power system results in an increase in $R_{G}$. On the other hand, $\mu_{N-1}$ and $\rho$ are the monotonic non-decreasing functions of the number of transmission lines in a network~\cite{Mieghem2011}. As a result, in Figure~\ref{fig:IEEE30MetricsVsThreshold}, removing the transmission lines from the system results in a decrease of $\mu_{N-1}$ and $\rho$. 

\subsection{Relating phase transitions to topology in power grids}
\label{subsec_Impact of line impedance}

To quantitatively assess the relationship between $l_{c}$ and the spectral graph metrics $R_{G}$, $\mu_{N-1}$, and $\rho$ the IEEE 30, IEEE 57, and the relatively larger IEEE 118 power systems are considered as use cases. These IEEE power systems are considered as reference systems, and an additional set of 99 synthetic power systems are generated for each power system following the methodology in Sec.~\ref{subsec_Test systems}. The topological changes in these test systems are quantified by $R_{G}$, $\mu_{N-1}$, and $\rho$. The test power systems are subjected to attacks, their robustness curves (See Figure~\ref{fig:ApprIEEE118RobCurve}) are determined, and the phase transition points under the electrical node significance and the betweenness based attacks are determined by $l_{c}$. For each test power system, the results obtained from the spectral radius, the algebraic connectivity, and the effective graph resistance are compared to the critical loading threshold of the systems to assess the relationship between these metrics.

Table~\ref{tab:IEEE30Linear}, Table~\ref{tab:IEEE57Linear}, and Table~\ref{tab:IEEE118Linear} quantitatively show to what extent $R_{G}$, $\rho$ and $\mu_{N-1}$ capture the phase transition behaviour in the IEEE 30, IEEE 57, and IEEE 118 power systems, respectively. It shows the typical values of Pearson's linear correlation coefficient~\cite{PearsonRef} $q$ between $l_{c}$ and the spectral graph metrics $R_{G}$, $\rho$, and $\mu_{N-1}$. The algebraic connectivity outperforms the other two metrics in estimating the phase transition behaviour of power grids. For example, in the IEEE 118 power system case, for both of the attack strategies, $\mu_{N-1}$ correlates positively to the critical loading point with a linear correlation coefficient of around 90\%, while the spectral radius has a linear correlation level of around 50\% (See Table~\ref{tab:IEEE118Linear}). 

\begin{table}[!htb]
\centering
\caption{Linear correlation coefficients $q$ between ($l_{c}$,$R_{G}$),
($l_{c}$,$\rho$), and ($l_{c}$,$\mu_{N-1}$) for the IEEE 30 power systems for $\delta$-based and $C_{B}$-based attacks.}
\label{tab:IEEE30Linear}
 \begin{tabular}{ l c c c c c c c c c } 
 \hline
 \hline
  &\multicolumn{1}{ c }{$\delta$-based attack} &\multicolumn{1}{ c }{$C_{B}$-based attack}\\
$q$						& 		$l_{c}$			&		$l_{c}$\\
 \hline
$R_{G}$ 				&		 -0.649		&		-0.662 \\

$\rho$ 				& 		 0.426 	  		&		0.5571\\ 

$\mu_{N-1}$ 		& 		 0.862 			& 		0.719\\ 
 \hline 
 \hline
 \end{tabular}
\end{table}

\begin{table}[!htb]
\centering
\caption{Linear correlation coefficients $q$ between ($l_{c}$,$R_{G}$),
($l_{c}$,$\rho$), and ($l_{c}$,$\mu_{N-1}$) for the IEEE 57 power systems for $\delta$-based and $C_{B}$-based attacks.}
\label{tab:IEEE57Linear}
 \begin{tabular}{ l c c c c c c c c c } 
 \hline
 \hline
  &\multicolumn{1}{ c }{$\delta$-based attack} &\multicolumn{1}{ c }{$C_{B}$-based attack}\\
$q$						& 		$l_{c}$			&		$l_{c}$\\
 \hline
$R_{G}$ 				&		 -0.739		&		-0.642 \\

$\rho$ 				& 		 0.722 	  		&		0.547\\ 

$\mu_{N-1}$ 		& 		 0.822 			& 		0.706\\ 
 \hline 
 \hline
 \end{tabular}
\end{table}

\begin{table}[!htb]
\centering
\caption{Linear correlation coefficients $q$ between ($l_{c}$,$R_{G}$),
($l_{c}$,$\rho$), and ($l_{c}$,$\mu_{N-1}$) for the IEEE 118 power systems for $\delta$-based and $C_{B}$-based attacks (See Figure\ref{fig:IEEE118NodeSignBased}).}
\label{tab:IEEE118Linear}
 \begin{tabular}{ l c c c c c c c c c } 
 \hline
 \hline
  &\multicolumn{1}{ c }{$\delta$-based attack} &\multicolumn{1}{ c }{$C_{B}$-based attack}\\
$q$						& 		$l_{c}$			&		$l_{c}$\\
 \hline
$R_{G}$ 				&		 -0.866		&		-0.834 \\

$\rho$ 				& 		 0.482 	  		&		0.576\\ 

$\mu_{N-1}$ 		& 		 0.921 			& 		0.932\\ 
 \hline 
 \hline
 \end{tabular}
\end{table}

The effective graph resistance has a negative correlation with the critical loading threshold while the spectral radius and algebraic connectivity correlate positively with the critical loading threshold. Increasing the impedance values of transmission lines of a power grid results in an increased (or unchanged) $R_{G}$ of the grid as $R_{G}$ is a monotonic non-decreasing function of the impedance values of the individual transmission lines~\cite{Ellens2011}. On the other hand, as the impedance values progressively increase, $\rho$ and $\mu_{N-1}$ of the system progressively decrease (or remain unchanged) because $\rho$ and $\mu_{N-1}$ are monotonic non-increasing functions of the impedance values of the individual transmission lines~\cite{Mieghem2011}. 

The spectral graph metrics are monotonic non-decreasing/-increasing functions of the impedance values of the individual transmission lines. Therefore, when creating the new test power systems, the progressive increase of impedance values results in sorted spectral graph metric values for these test power systems: $R_{G}$ in ascending order, $\mu_{N-1}$ and $\rho$ in descending order. Consequently, the Spearman's correlation coefficient that measures the strength of association between two \textit{ranked} variables, produces the same (absolute) result for ($l_{c}$,$R_{G}$), ($l_{c}$,$\rho$), and ($l_{c}$,$\mu_{N-1}$). Table~\ref{tab:IEEESpearman} shows the absolute values of the Spearman's correlation coefficient for the  IEEE 30, IEEE 57, and IEEE 118 power systems under $\delta$-based and $C_{B}$-based attacks. 

\begin{table}[!htb]
\centering
\caption{The absolute values of the Spearman's correlation coefficients $r_{s}$ between ($l_{c}$,$R_{G}$), ($l_{c}$,$\rho$), and ($l_{c}$,$\mu_{N-1}$) for the IEEE power systems for $\delta$-based and $C_{B}$-based attacks. $r_{s}$ is negative for ($l_{c}$,$R_{G}$), and positive for
($l_{c}$,$\rho$) and ($l_{c}$,$\mu_{N-1}$)}
\label{tab:IEEESpearman}
 \begin{tabular}{ l c c c c c c c c c } 
 \hline
 \hline

$r_{s}$						& 		$\delta$-based attack		&		$C_{B}$-based attack\\
 \hline
IEEE 30 				&		 0.667			&		0.691 \\

IEEE 57 				& 		 0.616 	  		&		0.630\\ 

IEEE 118 				& 		 0.845			& 		0.872\\ 
 \hline 
 \hline
 \end{tabular}
\end{table}

\begin{figure}[!htb]
\centering
\includegraphics[scale=0.5]{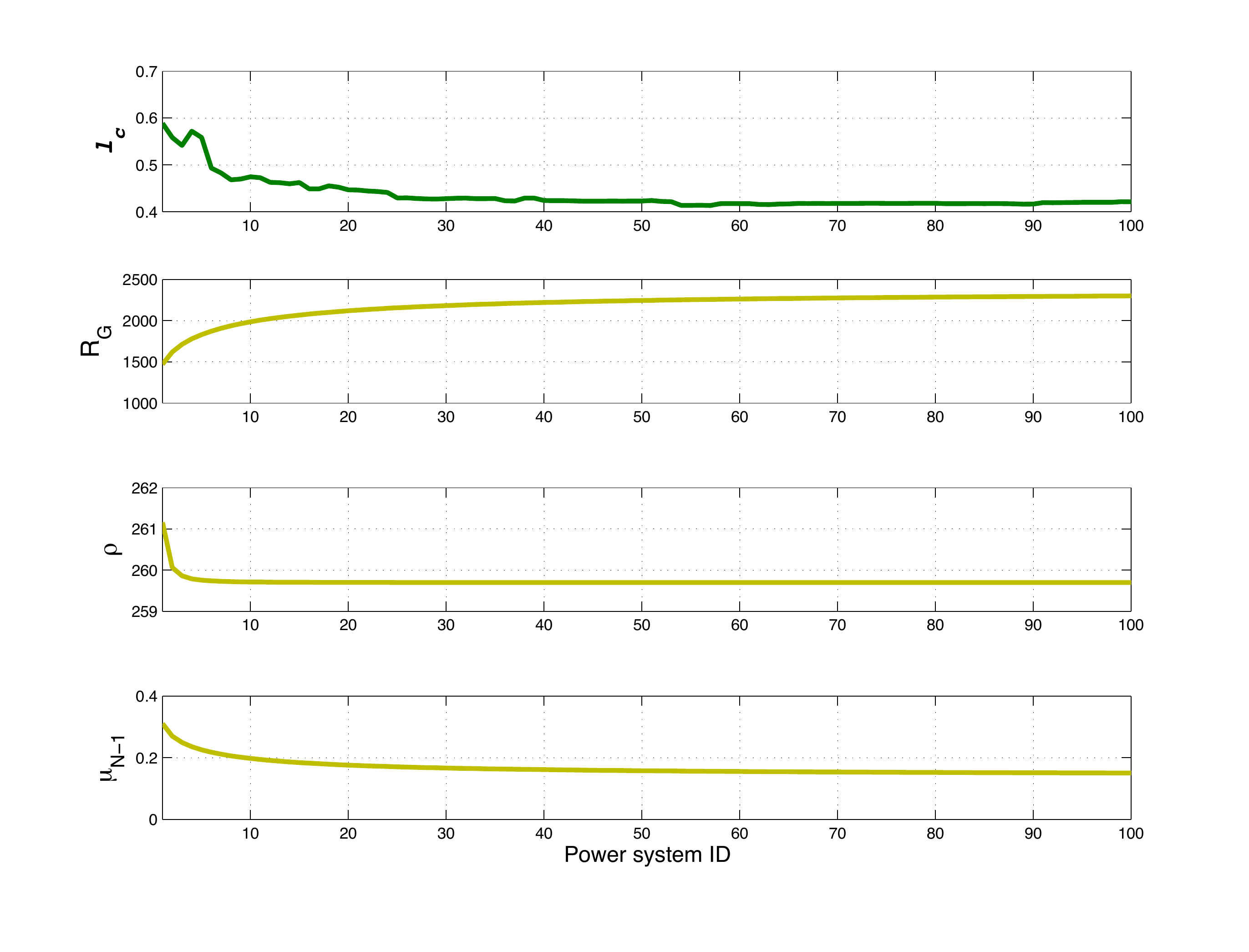}
\caption{Critical loading threshold and spectral graph metrics for the test power systems (consisting of the IEEE 118 power system and its derivative synthetic power systems). Due to differences in the impedance values of these systems, each test system has different values for the critical loading threshold, and for the spectral graph metrics. Different test systems are represented by 'Power system ID' (ranging from 1 to 100). Test systems are attacked with an electrical node significance based strategy.}
\label{fig:IEEE118NodeSignBased}
\end{figure}

\begin{figure}[!htb]
\centering
\includegraphics[scale=0.5]{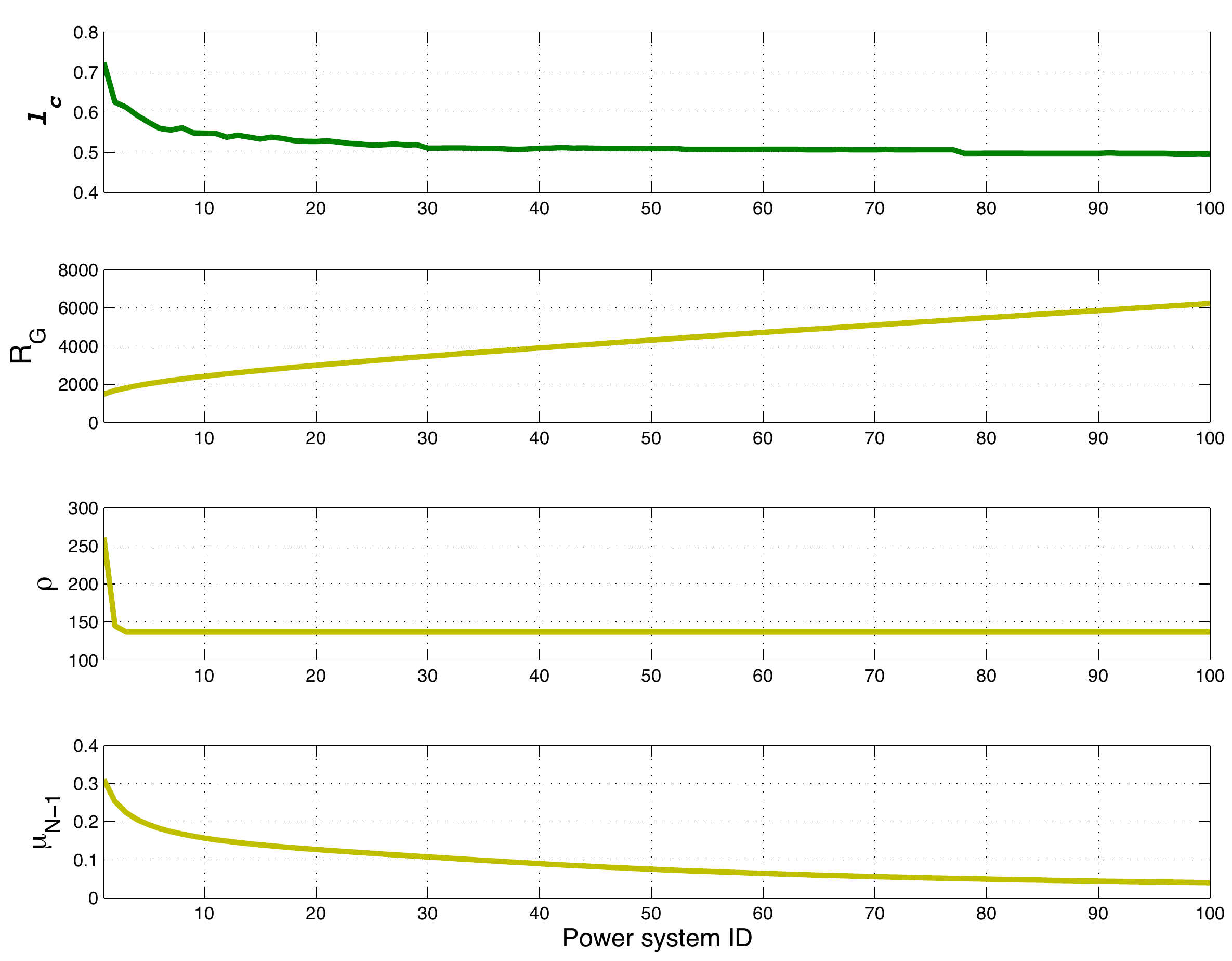}
\caption{Critical loading threshold and spectral graph metrics for the IEEE 118 and its derivative synthetic power systems. Test systems are attacked with a betweenness centrality based strategy.}
\label{fig:IEEE118BetwBased}
\end{figure}

Figure\ref{fig:IEEE118NodeSignBased} and Figure\ref{fig:IEEE118BetwBased} \textit{illustrates} the relationship between the metrics by plotting the critical loading threshold, effective graph resistance, spectral radius and algebraic connectivity for the test power systems obtained from the IEEE 118 power system. The test power systems are attacked with an electrical node significance $\delta$ (see Figure\ref{fig:IEEE118NodeSignBased}) and betweenness centrality $C_{B}$(see Figure\ref{fig:IEEE118BetwBased}) based strategy.

In both of the attack strategies in Figure\ref{fig:IEEE118NodeSignBased} and Figure\ref{fig:IEEE118BetwBased}, $l_{c}$ exhibits first a steep and then a slight decrease. As the impedance values of the chosen transmission lines increases, the ability of the grid to better distribute and accommodate the excess power flow over the grid degenerates. This results in an earlier phase transition in the damage due to cascading line overloads. All of the three metrics of effective graph resistance, spectral radius and algebraic connectivity capture this behaviour of the network and track the shift in the phase transition to a varying degree.

\begin{figure}[!htb]
\centering
\includegraphics[scale=0.45]{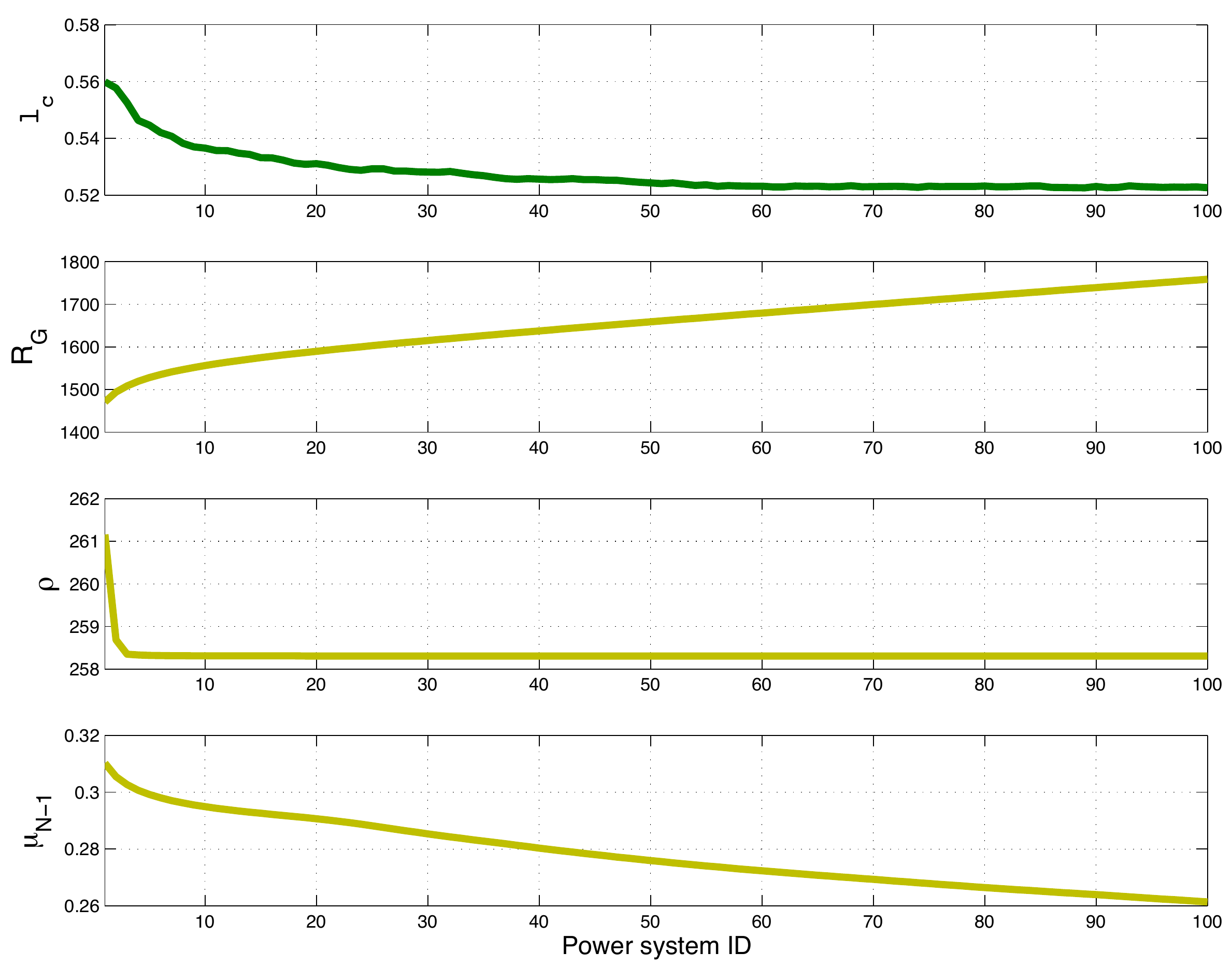}
\caption{Critical loading threshold and spectral graph metrics for the IEEE 118 and its derivative synthetic power systems. The results are averaged over 100 random instances. Test systems are attacked with an electrical node significance based strategy.}
\label{fig:IEEE118AveragedBy100CasesNodeSignBased}
\end{figure}

\begin{figure}[!htb]
\centering
\includegraphics[scale=0.45]{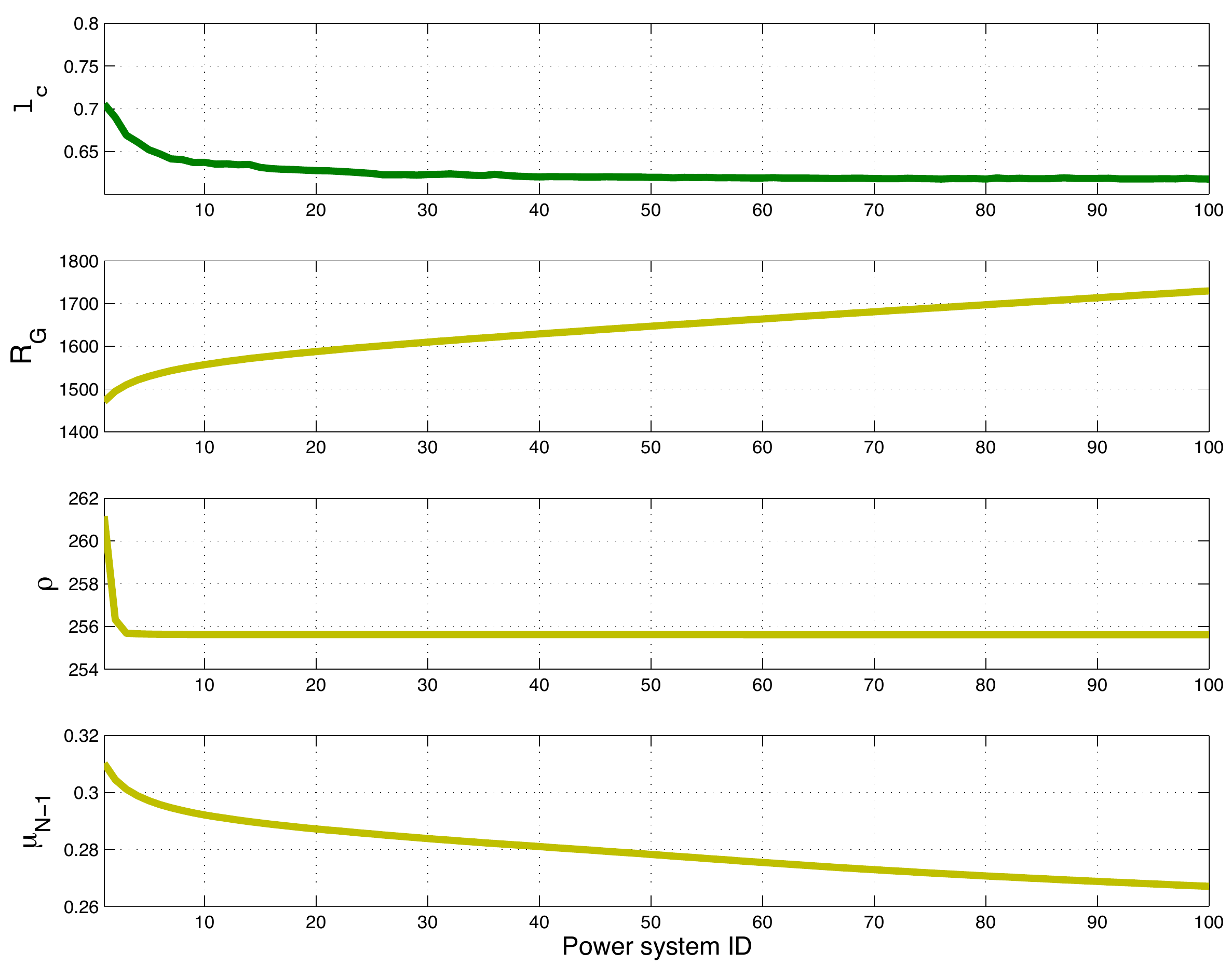}
\caption{Critical loading threshold and spectral graph metrics for the IEEE 118 and its derivative synthetic power systems. The results are averaged over 100 random instances. Test systems are attacked with an betweenness centrality based strategy.}
\label{fig:IEEE118AveragedBy100CasesBetweennessBased}
\end{figure}

To assess whether the spectral graph metrics capture the phase transition behaviour of power grids under different random transmission line choices when creating the synthetic power systems (See Sec.~\ref{subsec_Test systems}), the numerical analysis is extended. The IEEE 118 power system is considered as a reference case. The 10 \% of it's transmission lines are randomly chosen, additional 99 synthetic power systems are generated, and $l_{c}$, $R_{G}$, $\mu_{N-1}$, and $\rho$ values are determined for this set of 100 test systems. This process is repeated 100 times: 100 sets of 100 test power systems are generated and for each set of test systems $l_{c}$, $R_{G}$, $\mu_{N-1}$, and $\rho$ values are determined. Then, two different experiments are done: (i) $l_{c}$, $R_{G}$, $\mu_{N-1}$, and $\rho$ values are averaged over these 100 instances and the correlation level between these averaged results are determined, (ii) for each individual set of test systems, the correlation levels ($l_{c}$,$R_{G}$), ($l_{c}$,$\rho$), and ($l_{c}$,$\mu_{N-1}$) are determined. Figure~\ref{fig:IEEE118AveragedBy100CasesNodeSignBased} and Figure~\ref{fig:IEEE118AveragedBy100CasesBetweennessBased} illustrates the averaged $l_{c}$, $R_{G}$, $\mu_{N-1}$, and $\rho$ values, while Table~\ref{tab:IEEE118LinearAveraged} shows the correlation levels between these averaged results from (i). Table~\ref{tab:IEEE118LinearInd} shows the mean and the standard deviation of the individual correlation levels from (ii).

\begin{table}[!htb]
\centering
\caption{Linear correlation coefficients $q$ between ($l_{c}$,$R_{G}$),
($l_{c}$,$\rho$), and ($l_{c}$,$\mu_{N-1}$) for the IEEE 118 power systems for $\delta$-based and $C_{B}$-based attacks. The $l_{c}$, $R_{G}$, $\rho$, and $\mu_{N-1}$ are averaged over 100 different instances.}
\label{tab:IEEE118LinearAveraged}
 \begin{tabular}{ l c c c c c c c c c } 
 \hline
 \hline
  &\multicolumn{1}{ c }{$\delta$-based attack} &\multicolumn{1}{ c }{$C_{B}$-based attack}\\
$q$						& 		$l_{c}$			&		$l_{c}$\\
 \hline
$R_{G}$ 				&		 -0.873		&		-0.799 \\

$\rho$ 				& 		 0.495 	  		&		0.628\\ 

$\mu_{N-1}$ 		& 		 0.888			& 		0.839\\ 
 \hline 
 \hline
 \end{tabular}
\end{table}

The results in Figure~\ref{fig:IEEE118AveragedBy100CasesNodeSignBased}, Figure~\ref{fig:IEEE118AveragedBy100CasesBetweennessBased}, and Table~\ref{tab:IEEE118LinearAveraged} show that averaging the results over 100 instances (in which the transmission lines are randomly chosen when creating the test systems) does not change the results: $\mu_{N-1}$ again outperforms the other two metrics in anticipating the phase transition point of the IEEE 118 power system. The linear correlation coefficient remains around the same level compared to the individual correlation level (See Table~\ref{tab:IEEE118Linear} and Table~\ref{tab:IEEE118LinearAveraged}): \% 88.8 for a $\delta$-based attack strategy and \% 83.9 for a $C_{B}$-based attack strategy. The Spearman's rank correlation further increases to \% 96.5 for a $\delta$-based attack strategy and \% 95.8 for a $C_{B}$-based attack strategy.     

\begin{table}[!htb]
\centering
\caption{The mean and the standard deviation of the 100 linear correlation coefficients $q$ between ($l_{c}$,$R_{G}$), ($l_{c}$,$\rho$), and ($l_{c}$,$\mu_{N-1}$) for the IEEE 118 power systems for $\delta$-based attacks.}
\label{tab:IEEE118LinearInd}
 \begin{tabular}{ l c c c c c c c c c } 
 \hline
 \hline
  &\multicolumn{1}{ c }{Mean} &\multicolumn{1}{ c }{std}\\
$q$						& 		$l_{c}$			&		$l_{c}$\\
 \hline
$R_{G}$ 				&		 - 0.642		&		0.261 \\

$\rho$ 				& 		 0.551 	  		&		0.240\\ 

$\mu_{N-1}$ 		& 		 0.729			& 		0.278\\ 
 \hline 
 \hline
 \end{tabular}
\end{table}


\section{Conclusions}
\label{sec_Conclusion}

This paper investigates the impact of the topology of a power grid on the phase transitions in its robustness to cascading line overloads. Three spectral graph metrics, effective graph resistance $R_{G}$, spectral radius $\rho$ and algebraic connectivity $\mu_{N-1}$, that are used in the emerging field of network science to measure the robustness of networks, are considered and the relationship between these metrics and the phase transition behaviour of a power grid is assessed.   

The IEEE 118 buses power system and their derivative synthetic grids are used as test cases. For each power system (i) $R_{G}$, (ii) $\rho$, and (iii) $\mu_{N-1}$ are evaluated, while their phase transition points are determined by (iv) the critical loading threshold $l_{c}$. Since $l_{c}$ is based on simulations and takes into account specific details of power grids, the results from this metric are considered as the reference. The results from the first three metrics are compared to the results from $l_{c}$ to assess to what extent these graph theoretical metrics capture the phase transition behaviour of power grids. The numerical analysis on a model of power grids show that $\mu_{N-1}$ performs best to capture the phase transition behaviour with a correlation level e.g. in case of IEEE 118 power system around 90\%  while $\rho$ relates relatively less accurately to the phase transitions with a correlation level around 50\%. 

The experimental results in this paper suggest that $\mu_{N-1}$ performs best in estimating the phase transition in the robustness of power grids while $\rho$ of a power grid relates less accurately to these transitions compared to $\mu_{N-1}$ and $R_{G}$. This seems to contrast with the results from other fields such as epidemic processes and network synchronisation~\cite{Wang2003, Mieghem2009, Restrepo2005} in which the critical breakdown threshold is related to the spectral radius $\rho$ of the network. The phase transitions in the robustness of power grids are connectivity-based, while those in the dynamic systems such as epidemic processes and network synchronisation are process-based (and thus determined by the differential equations, whose solutions depend on the spectrum of the adjacency matrix of the underlying network). Consequently, the spectral radius relates to the process-based phase transitions (e.g. phase transitions in the epidemic processes), while the algebraic connectivity and the effective graph resistance that measure the connectivity of a network, relate to the connectivity-based phase transitions.    

The critical loading threshold $l_{c}$ defines a reference point for the system loading level at which the risk and the size of damage due to cascade increases abruptly. Designing and optimising the power system topologies appropriately with respect to this critical point enhances the robustness of power grids to cascading failures. The experimental results from this paper demonstrate the potential of the spectral graph metrics, in particular the potential of $\mu_{N-1}$, as measures to design/optimize power grid topologies for an enhanced phase transition behaviour.

The experimental results on the IEEE 118 power system (See Figure~\ref{fig:ApprIEEE118RobCurve}) show that increasing the loading level in a power grid may occasionally cause a higher level of robustness. This result suggests the existence of Brasess's Paradox in the electrical power systems. Braess's Paradox that is shown to occur in the road traffic networks, states that adding/creating extra capacity to/in a network can occasionally reduce the overall performance of a network. The future work will focus on the investigation of Braess's Paradox in electrical power grids. 

\subsection*{{\bf Acknowledgements}}

{\small
 This work is funded by the NWO project \emph{RobuSmart: Increasing
   the Robustness of Smart Grids through distributed energy
   generation: a complex network approach}, grant number
 647.000.001. The authors would like to thank the anonymous reviewer for his/her valuable comments.}



\vspace*{\fill}

\end{document}